\documentclass[fleqn,twoside]{article}
\usepackage{epsfig}
\usepackage{espcrc2}

\usepackage{graphicx}
\usepackage[figuresright]{rotating}

\newcommand{\AmS}{{\protect\the\textfont2
  A\kern-.1667em\lower.5ex\hbox{M}\kern-.125emS}}

\title{Ultra High Energy $\nu_\tau$ detection at Pierre Auger Observatory}

\author{G. Miele\address[DSF]{Dipartimento di Scienze Fisiche, Universit\`{a} di Napoli
``{\it Federico II} '' and  Istituto Nazionale di Fisica Nucleare
-- Sezione di Napoli, Complesso Universitario di Monte S. Angelo,
Via Cinthia, I-80126 Napoli, Italy.},
        L. Perrone\address{Fachbereich C, Sektion Physik, Universit\"{a}t Wuppertal,
D-42097 Wuppertal, Germany.},
        O. Pisanti\addressmark[DSF]}

\begin{document}

\begin{abstract}
Earth--skimming UHE tau neutrinos have a chance to be detected by
the Fluorescence Detector (FD) of Pierre Auger Observatory if
their astrophysical flux is large enough. A detailed evaluation of
the expected number of events is here performed for a wide class
of neutrino flux models. \vspace{1pc}
\end{abstract}

\maketitle

Neutrinos with energy above $10^{17}$ eV are expected to originate
from the interaction of UHE cosmic rays with the Cosmic Microwave
Background (CMB) {\it via} the $\pi$-photoproduction, $p +
\gamma_{CMB} \rightarrow n + \pi^+$, the so-called {\it cosmogenic
neutrinos} \cite{cosmogenic}. The prediction for such a flux have
been exhaustively discussed in several papers (see for example
Ref.s \cite{Kalashev:2002kx,Semikoz:2003wv}).

High energy neutrinos are hardly detected, as they are almost
completely shadowed by Earth and rarely interact with the
atmosphere. In this framework, an interesting strategy for
$\nu_\tau$ detection has been proposed in literature (see
\cite{Aramo:2004pr} for a complete list of references). For energy
between $10^{18}$ and $10^{21}$ eV the $\tau$ decay length is not
much larger than the corresponding interaction range. Thus, an
energetic $\tau$, produced by Charged Current (CC) $\nu_\tau$
interaction not too deep under the surface of the Earth, has a
chance to emerge in the atmosphere as an upgoing particle. Unlike
$\tau$'s, muons crossing the rock rapidly loose energy and decay.
Almost horizontal $\nu_\tau$, just skimming the Earth surface,
will cross an amount of rock of the order of their interaction
length and thus will be able to produce a corresponding $\tau$,
which might shower in the atmosphere and be detected.

A detailed estimate of the number of possible upgoing $\tau$
showers which the Fluorescence Detector (FD) of Pierre Auger Observatory
 could detect
is presented in Ref. \cite{Aramo:2004pr}, where the predictions
are analyzed with respect to their dependence on different
neutrino fluxes and by using a new estimate of neutrino-nucleon cross
sections. In the present paper a brief review of the main results
obtained in Ref. \cite{Aramo:2004pr} is reported.

Ultra High Energy protons, with energy above $\sim 10^{20}$ eV,
travelling through the universe mostly loose their energy {\it
via} the interaction with CMB radiation. The large amount of
charged and neutral pions produced will eventually decay in
charged leptons, neutrinos ({\it cosmogenic neutrinos}
\cite{cosmogenic}), and high energy gamma rays.

\begin{figure}[h]
\begin{center}
\epsfig{figure=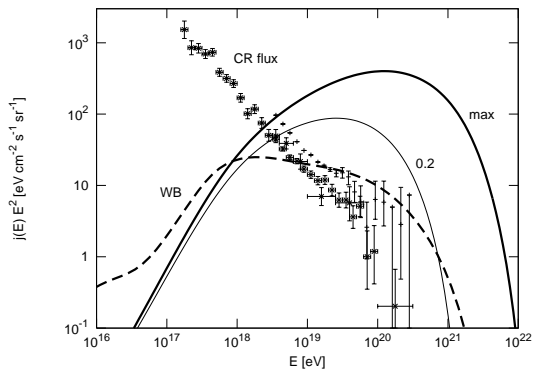,height=5cm} \caption{Cosmogenic
neutrino fluxes as a function of energy.} \label{nu_gzk}
\end{center}
\end{figure}
At the GeV energy range the extragalactic diffuse gamma-ray
background was measured by the EGRET experiment. This measurement
provides an upper bound for possible neutrino fluxes from pion
production. In particular, it gives the expected maximum flux of
cosmogenic neutrinos from an initial spectrum of measured UHE
protons. It is worth noticing that, since at least part of UHECR
are protons, the existence of cosmogenic neutrinos is guaranteed,
even if their flux is very uncertain. In Figure \ref{nu_gzk} the
GZK neutrino flux for three possible scenarios is plotted. The
thick solid line gives the case of an initial proton flux $\propto
1/E$, by assuming in addition that the EGRET flux is entirely due
to $\pi$-photoproduction (GZK-H). The thin solid line shows the
neutrino flux when the associated photons contribute only up to
20\% in the EGRET flux (GZK-L). The dashed line stands for the
conservative scenario of an initial proton flux $\propto 1/E^2$
(GZK-WB). In this case the neutrino flux is compatible with the
so--called Waxman-Bahcall limit.
\begin{figure}[h]
\begin{center}
\epsfig{figure=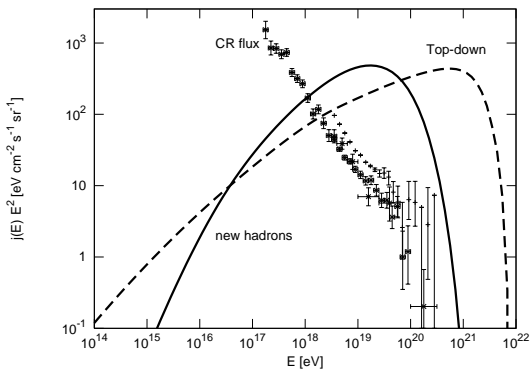,height=5cm} \caption{Neutrino
fluxes in exotic UHECR models.} \label{nu_exotic}
\end{center}
\end{figure}

Most of the models trying to explain highest energy cosmic rays
($E>10^{20}$ eV) in terms of exotic particles, predict a large
associated flux of neutrinos. In Figure \ref{nu_exotic}, the
expected neutrino flux for two of such scenarios is plotted. One
of them is the model of {\it new hadrons} (NH), with mass $M\sim
2-5$ GeV, capable of generating UHECR events above GZK cutoff. In
SUSY theories, for example, the new hadrons are bound states of
light bottom squarks or gluinos and, once produced in suitable
astrophysical environments, can reach the Earth without
significant energy losses. In spite the production of new hadrons
is a subdominant process, it generates a large number of neutrinos
(see Figure \ref{nu_exotic}).

The dashed line in Figure \ref{nu_exotic} shows the neutrino flux
for a Topological Defects model (TD) (for a review see
\cite{sigl_review}). In this case UHECR events with energy
$E>10^{20}$ eV are explained in terms of $\gamma$'s which are
produced in the decay of heavy particles with mass of the order of
$10^{22-23}$ eV. As in the previous case, the associated neutrino
flux for this kind of models is extremely large.

At energy above 1 GeV the interaction between neutrinos and the
atoms of the rock is dominated by the process of {\it Deep
Inelastic Scattering} (DIS) on nucleons. Detectable leptons are
produced through Charged Current interaction $\nu_{l}
(\bar{\nu}_{l}) + N \rightarrow l^{-} (l^{+}) + X $. The total
cross sections can be written in terms of differential ones as
follows
\begin{eqnarray}
\sigma_{CC}^{\nu N}(E_{\nu})&=&
\int_{0}^{1-\frac{m_{\l}}{E_{\nu}}}
              \frac{d\sigma_{CC}^{\nu N}}{dy} \,(E_{\nu},y) \, dy \,\,\, ,
\end{eqnarray}
where  $E_{\nu}$ is the energy of the incoming neutrino, $m_{\l}$
is the mass of the outgoing charged lepton and $y$ is the
inelasticity parameter, defined as
\begin{equation}
y_{CC}=1-\frac{E_{\l}}{E_{\nu}}\,\,\, ,
\end{equation}
with $E_{\l}$  the energy of the outgoing charged (for CC) or
neutral (for NC) lepton. In the present analysis we have used the
CTEQ6~\cite{cteq6} parton distribution functions in the DIS
factorization scheme. The $Q^2$--evolution is realized by the
next-to-leading order Dokshitzer-Gribov-Lipatov-Altarelli-Parisi
equations.

\begin{figure}[htb]
\begin{center}
\epsfig{figure=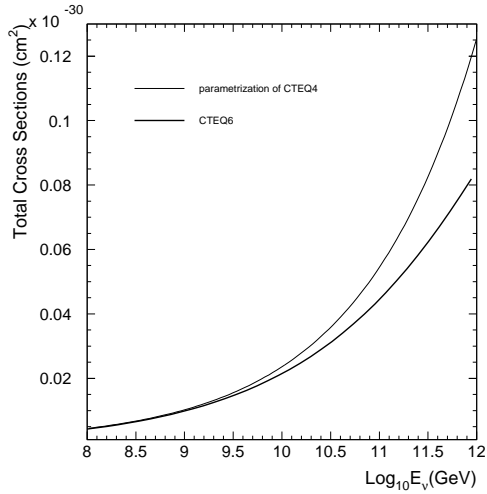,height=7cm} \caption{The total
$\nu_\tau$-nucleon CC cross sections based on both CTEQ4
\cite{quiggcross} and CTEQ6~\cite{cteq6} are reported.}
\label{comp}
\end{center}
\end{figure}

Figure~\ref{comp} shows a comparison between a CTEQ4-based
parametrization of CC cross section and the corresponding
calculation performed with CTEQ6. A substantial agreement is found
up to 10$^9$ GeV, whereas a discrepancy of at most 30\%  at
10$^{12}$ GeV is observed (CTEQ4 prediction being larger). In this
range of energy the uncertainty due to the lack of knowledge of
parton distribution functions is expected to be overwhelming.

\begin{figure}[htb]
\begin{center}
\epsfig{figure=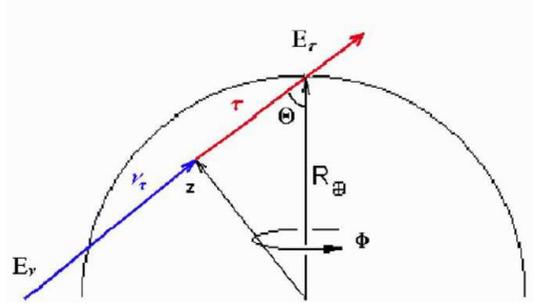,height=4cm} \caption{A neutrino
$\nu_\tau$ crosses the Earth with energy $E_\nu$ at a nadir angle
$\theta$ and azimuth angle $\phi$. Then it travels a distance $z$
before converting into a charged lepton $\tau$, which emerges from
the Earth surface with energy $E_\tau$ \cite{Feng:2001ue}.}
\label{geo1}
\end{center}
\end{figure}

Following the formalism developed in Ref. \cite{Feng:2001ue}, let
$\Phi_\nu$ be an isotropic flux of $\nu_\tau+\overline{\nu}_\tau$.
The differential flux of charged leptons emerging from the Earth
surface with energy $E_\tau$ is given by
\begin{eqnarray}
\frac{d\Phi_\tau(E_\tau,\Omega)}{dE_\tau\,d\Omega}&=& \int dE_\nu
\, \frac{d\Phi_\nu(E_\nu,\Omega)}{dE_\nu\,d\Omega} \nonumber\\
&\times & K(E_\nu,\,\Omega;\,E_\tau)\,\,\, , \label{eq:1}
\end{eqnarray}
where $K(E_\nu,\,\theta;\,E_\tau)$ is the probability that an
incoming neutrino crossing the Earth with energy $E_\nu$ and nadir
angle $\theta$ produces a lepton emerging with energy $E_\tau$
(see Figure~\ref{geo1}). In Eq.(\ref{eq:1}), due to the very high
energy of $\nu_\tau$, we can assume that in the process $\nu_\tau
\, + \, N \rightarrow \tau \, + \, X$ the charged lepton is
produced along the neutrino direction.
This process can occur if and only if the following conditions are
fulfilled:\\
a) the $\nu_\tau$ with energy $E_\nu$ has to survive along a
distance $z$ through the Earth;\\
b) the neutrino converts into a $\tau$ in the
interval $z, z+dz$;\\
c) the created lepton emerges from the Earth before decaying.\\

By using the expressions for all the above probabilities one
obtains
\begin{eqnarray}
&&K(E_\nu,\,\theta;\,E_\tau)= \frac{\sigma_{CC}^{\nu N}(E_\nu) \,
N_A}{E_\tau (\beta_\tau+\gamma_\tau \, E_\tau)} \,
\left(F(E_\nu,E_\tau)\right)^{\xi} \,
\nonumber\\
&&{\times} \exp\left\{-\frac{m_\tau}{c \tau_\tau \beta_\tau
\varrho_s}\left(\frac{1}{E_\tau}-
\frac{1}{E_\tau^0(E_\nu)}\right)\right. \nonumber\\
&& \left. -2 R_\oplus \cos{\theta} \, \sigma_{CC}^{\nu N}(E_\nu)
\, \varrho_s \, N_A \right\}\,\,\, , \label{eq:13}
\end{eqnarray}
where $\xi \equiv \omega + \sigma_{CC}^{\nu N}(E_\nu) \,
N_A/\beta_\tau$. As extensively discussed in Ref.
\cite{Aramo:2004pr} the parameters $\beta_\tau \simeq 0.71 {\cdot}
10^{-6}$ cm$^2$ g$^{-1}$ and $\gamma_\tau \simeq 0.35 {\cdot}
10^{-18}$ cm$^2$ g$^{-1}$ GeV$^{-1}$ fairly describe the $\tau$
energy loss in matter. Moreover, the tau lepton, once produced,
carries an average energy $E^0_{\tau}(E_\nu)=(1-<y_{CC}>)E_\nu$.

Eq.~(\ref{eq:13})
leads to the total rate of upgoing
$\tau$'s showering on the Auger detector, and thus {\it
potentially} detectable by the FD:
\begin{eqnarray}
\frac{d\textsl{N}_\tau}{dt}=2 \pi S \, D
\,\int_{E_\nu^{min}}^{E_\nu^{max} } dE_\nu
\int_{E_\tau^{th}}^{E_\tau^0(E_\nu)} dE_\tau \, \nonumber\\
\times \int_{\cos\theta_{min}}^{1}\,
\frac{d\Phi_\nu(E_\nu)}{dE_\nu\,d\Omega} \,
 K(E_\nu,\,\theta;\,E_\tau) \, \nonumber\\ \times \left(1-\exp\left\{-\frac{H \, m_\tau}
{c\tau_\tau \, E_\tau}\right\}\right) \, \varepsilon \, \cos\theta
\,d(\cos\theta)  \,\,\,, \label{eq:16}
\end{eqnarray}
where we have used the isotropy of the considered neutrino flux.
In Eq.(\ref{eq:16}) the quantity $S=3000\,\mathrm{km}^2$ is the
geometrical area covered by the Auger apparatus, $D\sim$ 10\%
 is the duty cycle for fluorescence detection, $ E_\tau^{th} \simeq
10^{18}$ eV is the energy threshold for the fluorescence process,
and $E_\nu^{min}$ is the minimum neutrino energy capable of
producing a $\tau$ at detection threshold. The quantity
$E_\nu^{max}$ is the endpoint of the neutrino flux.

The exponential term in the r.h.s. of Eq.(\ref{eq:16}) accounts
for the decay probability of a $\tau$ (showering probability) in a
distance $H$ from the emerging point on the Earth surface.

In Eq.(\ref{eq:16}) the integration over $\cos\theta$ can be
easily performed and this yields to the definition of the {\it
effective aperture} of the apparatus, $A(E_\nu)$, as
\begin{eqnarray}
\frac{d\textsl{N}_\tau}{dt}= D \, \int_{E_\nu^{min}}^{E_\nu^{max}
} dE_\nu \, \frac{d\Phi_\nu(E_\nu)}{dE_\nu\,d\Omega} \,
A(E_\nu)\,\,\, . \label{eq:17}
\end{eqnarray}
\begin{figure}
\begin{center}
\epsfig{figure=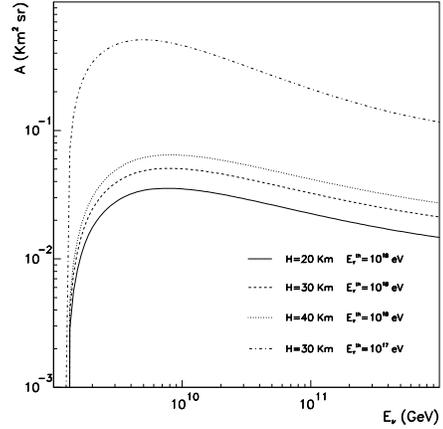,height=6cm} \caption{The {\it
effective aperture} $A(E_\nu)$ is here plotted for
$E^{th}_\tau=10^{17}$ eV and $H=30$ km, and for
$E^{th}_\tau=10^{18}$ eV with $H=20$, 30, and 40 km,
respectively.} \label{fig-aperture}
\end{center}
\end{figure}
In Figure \ref{fig-aperture} the quantity $A(E_\nu)$ is plotted
versus the neutrino energy for different values of $E^{th}_\tau$
and $H$.

The number of $\tau$-shower Earth--skimming events expected per
year at the FD detector for the different neutrino fluxes result
to be: 0.02 (GZK-WB), 0.04 (GZK-L), 0.09 (GZK-H), 0.11 (TD), 0.25
(NH).

As well known, neutrino induced extensive air showers (see Ref.
\cite{Ambrosio:2003nr} for an extended discussion) can be
disentangled by the ordinary cosmic ray background only for very
inclined and deep showers. An estimate of the expected number of
downgoing events within 30 km from the FD detector results to be
comparable with the upgoing one only for zenith angles larger than
$70^\circ$.

\end{document}